# Neutron-proton asymmetry dependence of spectroscopic factors in Ar isotopes

Jenny Lee(李曉菁)[1], M.B. Tsang(曾敏兒)[1], D. Bazin[1], D. Coupland[1], V. Henzl[1], D. Henzlova[1], M. Kilburn[1], W.G. Lynch(連致標)[1], A. Rogers[1], A. Sanetullaev[1], A. Signoracci[1], Z.Y. Sun(孙志宇)[1,2], M. Youngs[1], K.Y. Chae[3], R.J. Charity[4], H.K. Cheung (張凱傑)[5], M. Famiano[6], S. Hudan[7], P. O'Malley[8], W.A. Peters[8], K. Schmitt[3], D. Shapira[9], L.G. Sobotka[4]

[1]*NSCL & Department of Physics and Astronomy, Michigan State University, East Lansing, MI 48864, USA*
[2]*Institute of Modern Physics, CAS, Lanzhou 730000, People's Republic of China*
[3]*Department of Physics and Astronomy, University of Tennessee, Knoxville, Tennessee 37996, USA*
[4]*Department of Chemistry, Washington University, St. Louis, MO 63130, USA*
[5]*Physics Department, Chinese University of Hong Kong, Shatin, Hong Kong, China*
[6]*Department of Physics, Western Michigan University, Kalamazoo, MI 49008, USA*
[7]*Department of Chemistry, Indiana University, Bloomington, IN 47405, USA*
[8]*Department of Physics and Astronomy, Rutgers University, Piscataway, New Jersey 08854-8019, USA*
[9]*Oak Ridge National Laboratory, Oak Ridge, TN 37831, USA*

## Abstract

Spectroscopic factors have been extracted for proton rich $^{34}$Ar and neutron rich $^{46}$Ar using the (p,d) neutron transfer reaction. The experimental results show little reduction of the ground state neutron spectroscopic factor of the proton rich nucleus $^{34}$Ar compared to that of $^{46}$Ar. The results suggest that correlations, which generally reduce such spectroscopic factors, do not depend strongly on the neutron-proton asymmetry of the nucleus in this isotopic region as was reported in knockout reactions. The present results are consistent with results from systematic studies of transfer reactions but inconsistent with the trends observed in knockout reaction measurements.

Nuclei, like other Fermi liquids, are profoundly influenced by the interactions and the correlations between their fermion (nucleon) constituents [1,2]. Short-range correlations between nucleons mitigate their strong short-range repulsion, modifying the compressibility of nuclei and of nuclear matter [1,3]. Long-range correlations between valence nucleons induce collective modes and pairing correlations within nuclei [1,4]. These influence the spectrum of excitations near the Fermi surface and modify the heat capacity, consequences well known in nuclei and other Fermi liquids such as liquid $^3$He. Pairing correlations, analogous to superconductivity and superfluidity in macroscopic systems [5], couple pairs of valence neutrons or protons in nuclei to spin zero.

The correlated interplay between the single-particle and collective dynamics spreads the contributions from single-particle orbits over a large range in excitation energy [1,6]. This alters the occupancies of these orbits, which can be probed by measuring nucleon spectroscopic factors that describe the contributions of specific single-particle orbits to specific states. These measurements indicate that correlations reduce valence orbital occupancies by about 30% [1,6,7]. Some of the correlations due to the long-range components of the interaction vary strongly from nucleus to nucleus and have been described within current shell models [8]. Other factors that govern the remaining long-range correlations have been identified and are being studied [9].

Measurements of single nucleon removal mechanisms, such as single nucleon transfer or knockout, provide constraints on spectroscopic factors (SF's) calculated from the overlap between many-body wave functions of the initial and final states [10]. Large reductions (up to 75%) in the measured SF values relative to shell model predictions have been observed in one-nucleon knockout reactions for strongly bound valence nucleons [11]. Such large reductions in SF's suggest qualitatively new phenomena may play a role in these two component (neutron and proton) Fermi liquids. Even though the reduced SF's of strongly bound minority constituents in these nuclei are not theoretically anticipated, some studies do predict smaller effects. For example, Dispersive-Optical-Model analyses which include the effects of long-range correlations, predict changes in the neutron SF magnitudes from $^{40}$Ca to $^{49}$Ca of about10% [12].

In this paper, we investigate these effects using an alternative spectroscopic probe: (p,d) single-nucleon transfer reactions. Recently, a consistent and systematic understanding of SF's from single neutron transfer was obtained for (p,d) and (d,p) reactions by comparing cross-section data to calculations based on the Adiabatic Distorted Wave Approximation (ADWA) and nucleon global elastic scattering optical potentials [13,14]. Using this approach with the Chapel-Hill (CH89) nucleon-nucleus global optical potential and a neutron potential with fixed radius and diffuseness parameters, $r_0$=1.25 and $a_0$=0.65 fm, agreement to within 20% was obtained between measured neutron SF's and those calculated via Large Basis-Shell Model (LB-SM) calculations for ground states of nuclei with 3≤Z≤28 [13,14]. For most excited states of stable nuclei with 3≤Z≤24, the agreement is slightly worse, but within 30% [14]. If one uses a different optical model potential, developed by Jeukenne, Lejeune, and Mahaux (JLM) with conventional scale factors of $\lambda_V$ = 1.0 and $\lambda_W$ = 0.8 for the real and imaginary parts, and constrains the geometry of these potentials and that of the transferred-neutron bound state by Hartee-Fock calculations, one observes an overall reduction ~30% in the measured ground state spectroscopic factors [15]. This implies reduction factors $Rs$ ≡ (experimental SF)/(theoretical SF) of 30% in the latter approach, similar to the reductions in proton SF's extracted from (e,e'p) measurements [16].

Regardless of the choice of optical model potential or the geometry of the mean field potential for the transferred neutron, systematic analyses of neutron transfer reactions display no strong dependence of the reduction factor $Rs$ on the neutron-proton asymmetry of the nuclei [13-15]. However, systematic uncertainties inherent in comparing SF's from different experiments published over a period of more than 40 years reduce the sensitivity of such studies.

The available data using stable beams and targets include very few extremely neutron-rich or neutron-deficient nuclei. To explore more extreme asymmetries, we extracted the ground-state neutron SF's for $^{34}$Ar and $^{46}$Ar from (p,d) reactions using proton rich $^{34}$Ar and neutron rich $^{46}$Ar beams in inverse kinematics. SF's from knockout reactions on these nuclei have been published and a significant reduction of the neutron SF for $^{34}$Ar has been reported. [11]. The difference between the neutron and proton separation energy (ΔS), which characterizes the relative shift of neutron and proton Fermi energies in these nuclei, is 12.41 MeV and -10.03 MeV for $^{34}$Ar and $^{46}$Ar respectively. In previous studies of transfer reactions, there were no nuclei with |ΔS| >> 7 MeV [13,15].

In the present experiments, the deuteron angular distributions from p($^{34}$Ar,d)$^{33}$Ar and p($^{46}$Ar,d)$^{45}$Ar transfer reactions were measured using radioactive secondary beams of $^{34}$Ar and $^{46}$Ar at 33 MeV/nucleon at the National Superconducting Cyclotron Laboratory at Michigan State University [17]. The p($^{36}$Ar,d)$^{35}$Ar reaction was also measured using a degraded $^{36}$Ar primary beam at 33MeV/nucleon to compare with data previously measured in normal kinematics [18]. These beams were transported and focused on CH$_2$ targets with areal densities of 7.10 mg/cm$^2$ for $^{34,36}$Ar and 2.29 mg/cm$^2$ for $^{46}$Ar reactions. Deuterons were detected in the High-Resolution Array (HiRA) [19] in coincidence with the recoil residues detected in the S800 focal plane [20]. An array of 16 HiRA telescopes [19] was placed at 35 cm from the target where they subtended polar angles of 6°≤ $\theta_{lab}$ ≤45°. Each telescope contained 65$\mu m$ thick ΔE and 1500 $\mu m$ thick E silicon strip detectors, backed by 3.9 cm thick CsI(Tl) crystals. The strips in these telescopes effectively subdivided each telescope into 1024 2$mm$ x 2$mm$ pixels. Detailed descriptions of experimental setup can be found in Ref. [17].

Deuterons were identified in HiRA with standard energy loss techniques using the energy deposited in the ΔE and E Silicon strip and CsI detectors. Reaction residues were identified in the S800 spectrometer using the energy loss and the time-of-flight (TOF) information of the focal plane detectors [20]. Fig.1 (a-c) show the Q value spectra for deuterons that stop in the thick Si detector for p($^{34, 36,46}$Ar,d)$^{33, 35, 45}$Ar. The observed resolutions of 500, 470 and 410 keV FWHM for the transitions to the ground-states of $^{33,35,45}$Ar respectively agree with the expectation from GEANT4 [21] simulations taking into account the finite beam spot size, the energy resolution of the Si detectors, energy loss and angular straggling in the target. Measurements using a 1.7 mg/cm$^2$ carbon target reveal that the background from reactions on carbon is negligible when both deuteron and the heavy recoil are detected. The absolute normalization of the cross section was achieved to within 10% by directly counting the beam particles with a Micro-Channel Plate detector [22] placed ~10 cm upstream of the target. This also provided the start TOF signal for particles detected by the S800 spectrometer.

Figs. 1(d), 1(e) and 1(f) show the differential cross sections for the ground-state transition of p($^{34}$Ar,d)$^{33}$Ar, p($^{36}$Ar,d)$^{35}$Ar, and p($^{46}$Ar,d)$^{45}$Ar, respectively. The solid circles in the lower panels denote the data from present measurements and the open squares in Fig. 1(e) denote previous $^{36}$Ar (p,d)$^{35}$Ar data in normal kinematics at 33.6 MeV/nucleon [18]. The agreement between the measured cross sections from the present work and Ref.[18] for the first excited state is also very good [17]. For p($^{46}$Ar,d)$^{45}$Ar reaction, the ground state ($f_{7/2}$) and the first excited state (542 keV, $p_{3/2}$) were not resolved for center-of-mass angles larger than 8°. Fortunately, the $l$ values for the ground state ($l=3$) and first excited state ($l=1$) are different, resulting in very different angular distributions. Specifically, the angular distribution for the excited state exhibits a deep minimum near $\theta_{c.m.}$ =20-27°, close to a factor of 100 smaller than that of ground state, therefore, the cross-sections for the ground state could be unambiguously extracted [17].

The dashed curves in Figs. 1(d-f) show the ADWA calculations using the CH89 potential with the conventional neutron bound state Woods Saxon potential parameter, $r_0$=1.25 and $a_0$=0.65fm. The solid lines in Figs. 1(d-f) show the ADWA calculations using the JLM microscopic potential and the bound-state neutron potential, which have been constrained by Hartree-Fock calculations. Both calculations reproduce the shape of experimental angular distributions. Normalizing the ADWA model calculations to the data results in the SF values listed in Table 1. Similar to previous analyses, SF(JLM+HF) values are about 30% smaller than the SF(CH89) values. The ground-state neutron SF's for $^{34}$Ar and $^{36}$Ar were calculated in the sd-shell model space using USDB effective interaction [23]. The ground-state neutron SF for $^{46}$Ar was calculated in the sd-pf model space using the interaction of Nummela et al. [24].

The shell model ground state SF values as well as the reduction factors $Rs$, defined as the ratio of the experimental SF value divided by the LB-SM prediction are listed in Table 1. In Fig. 2, $Rs$ are plotted as a function of the difference in the neutron and proton separation energies, $\Delta S$, as open circles ($Rs$(CH89) = SF(CH89)/SF(LB-SM)) and closed circles ($Rs$(JLM+HF) = SF(JLM+HF)/SF(LB-SM)). The error bars listed in Table 1 and associated with both open and close circles in Fig. 2 reflect the uncertainties in the absolute cross-section determination [17]. Consistent with previous systematic studies with stable nuclei [13-15], the values of $Rs$ for symmetric $^{36}$Ar and neutron-rich $^{46}$Ar are similar. The extracted value of $Rs$ for proton-rich $^{34}$Ar is about 15-20% smaller. With the experimental uncertainties of $\pm$10% for both $^{34}$Ar and $^{46}$Ar, reductions in the spectroscopic factors for $^{34}$Ar relative to neutron rich $^{46}$Ar of 0-35% are possible but much larger reductions are excluded.

The weak dependence of reduction factors on the asymmetry of the three Ar isotopes is similar to the trends obtained from the recent Dispersive-Optical-Model analysis of elastic-scattering and bound-level data for $^{40-49}$Ca isotopes [12]. In contrast, a much larger systematic suppression in SFs has been reported for the knockout reactions [11], when the removed nucleon has a large separation energy or asymmetry. As shown by the open triangles in Fig. 2, the neutron $Rs$ extracted from knockout reactions for $^{34}$Ar is approximately a factor of two smaller than that for $^{46}$Ar [11]. Even larger reductions have been observed for neutron knockout from $^{32}$Ar [11], a nucleus for which transfer data is not available. This suggests that there is a systematic difference between the conclusions drawn from these two probes for the spectroscopic factors of strongly bound particles. Thus a reexamination of the reaction theory description of transfer reactions or knockout reactions including the input parameters used in these analyses may be needed.

In summary, we have extracted the neutron ground-state spectroscopic factors of $^{46}$Ar, $^{36}$Ar, $^{34}$Ar using (p,d) transfer reactions with radioactive beams in inverse kinematics. The experimental results, analyzed with two different approaches using different optical model potentials and different neutron bound-state geometries, are consistent with extensive systematics of spectroscopic factors obtained from transfer reactions on stable nuclei. Unlike the trends observed for knockout reactions, comparison of the extracted spectroscopic factors for proton-rich $^{34}$Ar and neutron rich $^{46}$Ar suggests a weak dependence of correlations on neutron-proton asymmetry in this isotope region. These new results examine the nature of nucleon correlations in nuclei with unusual isospin asymmetries while posing questions about the reaction mechanisms of transfer and knockout reactions used to probe them.

Acknowledgement

The authors would like to thank Professors B.A. Brown and J. Tostevin for the use of the programs Oxbash and TWOFNR. This work is supported by the National Science Foundation under grants PHY-0606007. Cheung acknowledges the support of the Summer Undergraduate Research Experience (SURE) program sponsored by the Chinese University of Hong Kong. Sun acknowledges support from the Chinese Academy of Sciences and MSU during his stay at NSCL in 2008.

[1] W. H. Dickhoff and D. V. Neck, *Many-body theory exposed*, World Scientific, Singapore (2008).
[2] D. Pines and P. Nozières, *The Theory of Quantum Liquids*, Benjamin, New York (1966).
[3] A. Akmal, V. R. Pandharipande, and D. G. Ravenhall, Phys. Rev. C 58, 1804 (1998).
[4] G. Golò, P.F. Bortignon, and R.A. Broglia, Nucl. Phys. A649, 335 (1999).
[5] M. A. Preston and R. K. Bhaduri, *Structure of the Nucleus*, Addison-Wesley Pub. Co. Boston (1974).
[6] W. H. Dickhoff and C. Barbieri, Prog. Part. Nucl. Phys. 52, 377 (2004).
[7] V. R. Pandharipande et al., Rev. Mod. Phys. 69, 981 (1997).
[8] B. A. Brown, Prog. Part. Nucl. Phys. 47, 517 (2001).
[9] C. Barbieri, Phys. Rev. Lett. 103, 202502 (2009).
[10] N. Austern, *Direct Nuclear Reaction Theories*, John Wiley & Sons, New York (1970).
[11] A. Gade et al., Phys. Rev. C77, 044306 (2008) and reference therein.
[12] R. J. Charity et al., Phys. Rev. C76, 044314 (2007).

[13] M. B. Tsang et al, Phys. Rev. Lett. 95, 222501 (2005) and reference therein.

[14] M. B. Tsang et al., Phys. Rev. Lett. 102, 062501 (2009).

[15] J. Lee et al, Phys. Rev. C73, 044608 (2006) and reference therein.

[16] G.J. Kramer et. al., Nucl. Phys. A. 679, 267 (2001) and references therein.

[17] J. Lee, PhD thesis, Michigan State University (2010).

[18] R. L. Kozub, Phys. Rev. 172, 1078 (1968).

[19] M. S. Wallace et al., Nucl. Instrum. Methods Phys. Res. A583, 302 (2007)

[20] D. Bazin et al., Nucl. Instrum. Methods Phys. Res. B 204, 629 (2003).

[21] S. Agostinelli et al., Nucl. Instrum. Meth. A, 506, 250 (2003).

[22] D. Shapira et al., Nucl. Inst. and Meth. A. 454, 409 (2000).

[23] B. A. Brown et al, Phys. Rev. C74, 034315 (2006).

[24] A. Signoracci and B. Alex Brown, Phys. Rev. Lett. 99, 099201 (2007).

[25] In this work, the ground state to ground state transition cross sections are determined but in Ref. [11], inclusive cross sections with contributions from the excited states to the ground states were measured. Thus the $^{34}$Ar $\Delta S$ value in Ref. [11] is weighted by the nucleon separation energy of the excited states. In principle, the knockout value should give the upper limit of the $^{34}$Ar ground-state $Rs$ value as contributions from excited states would increase the reduction factor.

Fig. 1. (Color online) Q-value spectrum (a-c, top panels) and deuteron angular distributions (d-f, bottom panels) to ground state of p($^{34,36,46}$Ar,d)$^{33, 35, 46}$Ar. The open squares in the bottom middle panel are data from previous normal kinematics experiments [18]. The solid and dashed lines represent the calculations using JLM+HF and CH89 approach respectively.

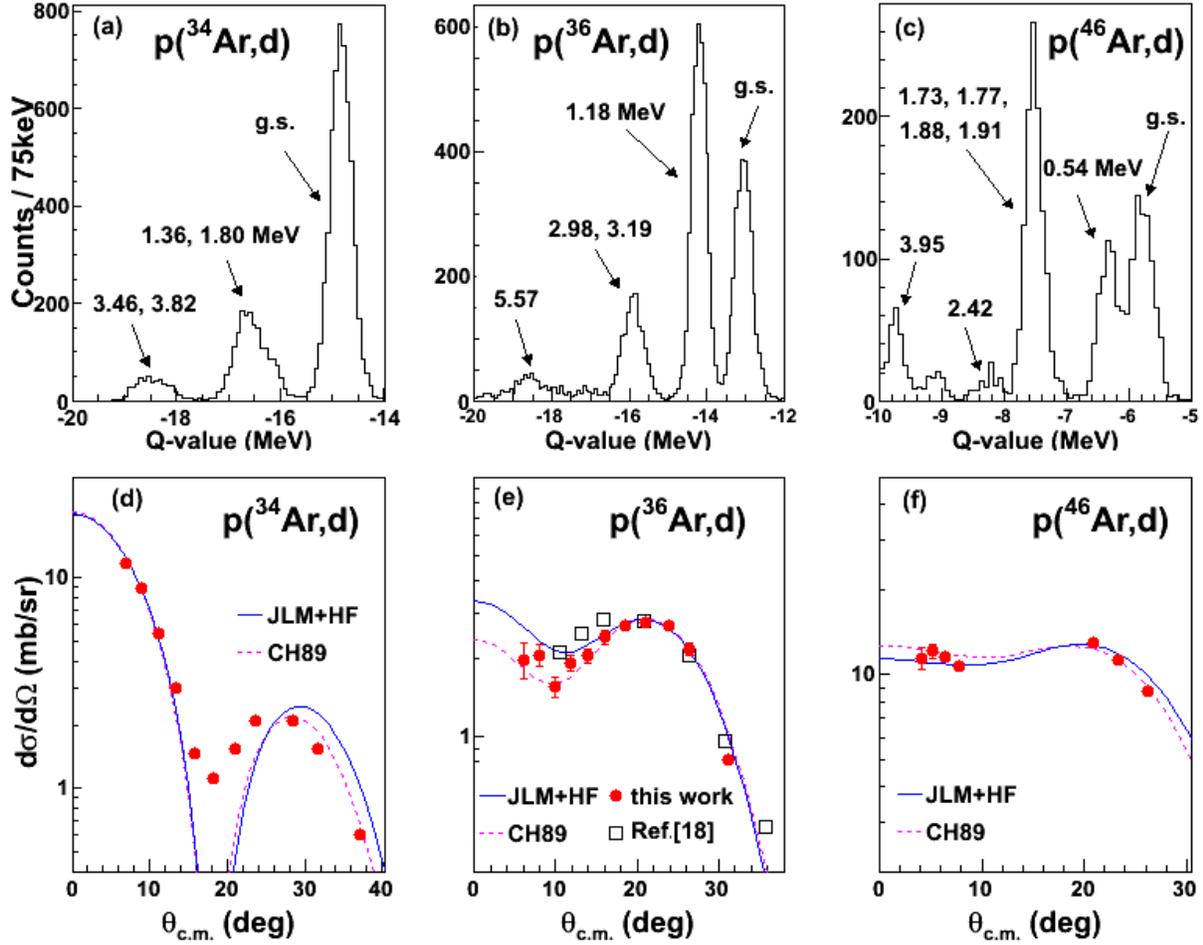

.

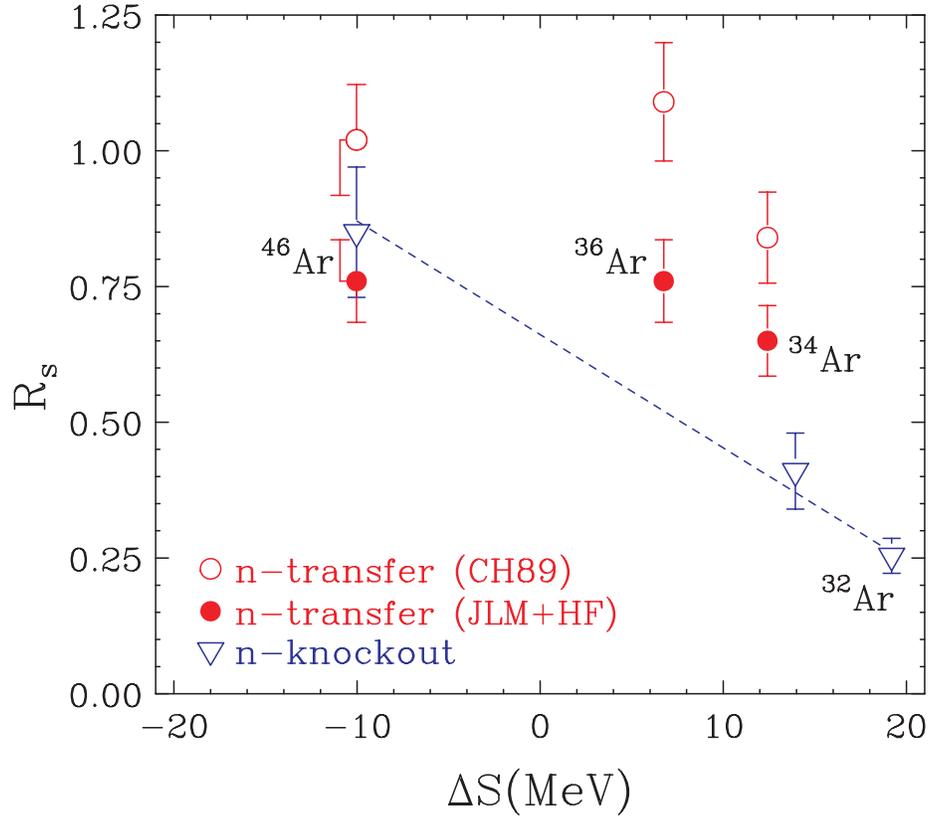

Fig. 2. (Color Online): Reduction factors $Rs$=SF(expt)/SF(LB-SM) as a function of the difference between neutron and proton separation energies, $\Delta S$. The solid and open circles represent $Rs$ deduced in JLM+HF and CH89 approach using the present transfer reaction data respectively. The open triangles denote the $Rs$ from knockout reactions [11]. The dashed line is the best fit of $Rs$ of $^{32,34,46}$Ar from knockout reactions. The use of different $\Delta S$ values from the present work and knockout reactions in Ref. [11] is explained in Ref. [25].

Table 1. Experimental and theoretical neutron spectroscopic factors (SF) and reduction factors ($Rs$) for ground state $^{34}$Ar, $^{36}$Ar and $^{46}$Ar.

| Isotopes | $lj^{\pi}$ | Sn(MeV) | $\Delta S$ (MeV) | (theo.) SF(LB-SM) | (expt.) SF(JLM+HF) | (expt.) Rs(JLM+HF) | (expt.) SF(CH89) | (expt.) Rs(CH89) |
|---|---|---|---|---|---|---|---|---|
| $^{34}$Ar | $s1/2^+$ | 17.07 | 12.41 | 1.31 | 0.85±0.09 | 0.65±0.07 | 1.10±0.11 | 0.84±0.08 |
| $^{36}$Ar | $d3/2^+$ | 15.25 | 6.75 | 2.10 | 1.60±0.16 | 0.76±0.08 | 2.29±0.23 | 1.09±0.11 |
| $^{46}$Ar | $f7/2^-$ | 8.07 | -10.03 | 5.16 | 3.93±0.39 | 0.76±0.08 | 5.29±0.53 | 1.02±0.10 |